\documentclass[10pt]{report}
\usepackage[top=0.85in,left=0.85in,footskip=0.75in]{geometry}
\usepackage{subcaption}
\usepackage{hyperref}
\usepackage{graphicx}

\pdfoutput=1

\usepackage{amsmath,amssymb}

\usepackage{nameref,hyperref}

\usepackage{microtype}
\DisableLigatures[f]{encoding = *, family = * }

\usepackage[table]{xcolor}

\usepackage{array}

\newcolumntype{+}{!{\vrule width 2pt}}

\newlength\savedwidth

\usepackage{setspace}
\raggedright
\makeatletter
\renewcommand{\@biblabel}[1]{\quad#1.}
\makeatother

\date{}

\begin{document}
%\vspace*{0.2in}

% Title must be 250 characters or less.
\begin{flushleft}
{\Large
\textbf\newline{Estimation of temporal covariances in pathogen dynamics using Bayesian multivariate autoregressive models } % Please use "sentence case" for title and headings (capitalize only the first word in a title (or heading), the first word in a subtitle (or subheading), and any proper nouns).
}
\newline
% Insert author names, affiliations and corresponding author email (do not include titles, positions, or degrees).
\\

Colette Mair\textsuperscript{1,2*},
Sema Nickbakhsh\textsuperscript{1},
Richard Reeve\textsuperscript{2},
Jim McMenamin\textsuperscript{3},
Arlene Reynolds\textsuperscript{3},
Rory N. Gunson\textsuperscript{4},
Pablo R. Murcia\textsuperscript{1},
Louise Matthews\textsuperscript{2}
\\
\bigskip
\textbf{1} MRC-University of Glasgow Centre for Virus Research, Institute of Infection, Immunity and Inflammation, College of Medical, Veterinary and Life Sciences, University of Glasgow, Glasgow, UK.
\\
\textbf{2} Boyd Orr Centre for Population and Ecosystem Health, Institute of Biodiversity, Animal Health and Comparative Medicine, College of Medical, Veterinary and Life Sciences, University of Glasgow, Glasgow, UK.
\\
\textbf{3} Health Protection Scotland, NHS National Services Scotland, Glasgow, UK.
\\
\textbf{4} West of Scotland Specialist Virology Centre, NHS Greater Glasgow and Clyde, Glasgow, UK.
\\
\bigskip

% Insert additional author notes using the symbols described below. Insert symbol callouts after author names as necessary.
%
% Remove or comment out the author notes below if they aren't used.
%
% Primary Equal Contribution Note
%\Yinyang These authors contributed equally to this work.

% Additional Equal Contribution Note
% Also use this double-dagger symbol for special authorship notes, such as senior authorship.
%\ddag These authors also contributed equally to this work.

% Current address notes
%\textcurrency Current Address: Dept/Program/Center, Institution Name, City, State, Country % change symbol to "\textcurrency a" if more than one current address note
% \textcurrency b Insert second current address
% \textcurrency c Insert third current address

% Deceased author note
%\dag Deceased

% Group/Consortium Author Note
%\textpilcrow Membership list can be found in the Acknowledgments section.

% Use the asterisk to denote corresponding authorship and provide email address in note below.
* Colette.Mair@glasgow.ac.uk

\end{flushleft}
% Please keep the abstract below 300 words
\section*{Abstract}
% * <colette.mair@glasgow.ac.uk> 2018-02-21T21:08:48.041Z:
%
% Example
%
%
% ^ <colette.mair@glasgow.ac.uk> 2018-02-21T21:08:59.933Z.
It is well recognised that animal and plant pathogens form complex ecological communities of interacting organisms within their hosts. Although community ecology approaches have been applied to determine pathogen interactions at the within-host scale, methodologies enabling robust inference of the epidemiological impact of pathogen interactions are lacking. Here we developed a novel statistical framework to identify statistical covariances from the infection time-series of multiple pathogens simultaneously. Our framework extends Bayesian multivariate disease mapping models to analyse multivariate time series data by accounting for within- and between-year dependencies in infection risk and incorporating a between-pathogen covariance matrix which we estimate. Importantly, our approach accounts for possible confounding drivers of temporal patterns in pathogen infection frequencies, enabling robust inference of pathogen-pathogen interactions. We illustrate the validity of our statistical framework using simulated data and applied it to diagnostic data available for five respiratory viruses co-circulating in a major urban population between 2005 and 2013: adenovirus, human coronavirus, human metapneumovirus, influenza B virus and respiratory syncytial virus. We found positive and negative covariances indicative of epidemiological interactions among specific virus pairs. This statistical framework enables a community ecology perspective to be applied to infectious disease epidemiology with important utility for public health planning and preparedness.

% Please keep the Author Summary between 150 and 200 words
% Use first person. PLOS ONE authors please skip this step.
% Author Summary not valid for PLOS ONE submissions.
\section*{Author summary}
Disease-causing microorganisms, including viruses, bacteria, protozoa and fungi form complex ecological communities within the animals and plants they inhabit. These microorganisms can coexist harmoniously or even beneficially, or they may competitively interact for host resources. Well-studied examples include interactions between viruses and bacteria in the gut and respiratory tract of mammals. Whilst previous experimental and epidemiological studies have revealed that some pathogens do interact within their hosts, demonstrating their epidemiological significance is challenging. This is in part due to a lack of large-scale data describing the infection patterns of multiple pathogens within single populations over long time frames. Methods for evaluating whether infection frequencies of different pathogens fluctuate together or not over time do not readily adjust for alternate explanations. For example, human pathogens may have similar or different seasonal patterns depending on the age groups they infect and the weather conditions they survive in, and not because they are directly or indirectly interacting. We developed a robust statistical framework that allows pathogen-pathogen interactions to be identified from population scale diagnostic data. This framework provides new insights into pathogen interactions and will have important consequences for public health preparedness and the design of effective disease control interventions.

%\linenumbers

% Use "Eq" instead of "Equation" for equation citations.
\section*{Introduction}
Animals and plants are exposed to a diverse community of pathogenic organisms that co-circulate in time and space. When multiple pathogens infect the same tissue, they effectively share an ecological niche that provides the opportunity for interspecific interactions \cite{Telfer2010, Rynkiewicz2015, Seabloom2015}. It is known that pathogen interactions may alter the within-host dynamics of infection with consequences for the population transmission of some common infections. Interactions among microorganisms include the promoting or inhibiting effects of gut microbiota on invading pathogenic bacteria in the gastrointestinal tract \cite{Baumler2016}; the enhanced carriage of pneumococcal bacteria following influenza infection in the respiratory tract \cite{Mina2014}; and immune-driven enhancement of Zika virus infection following Dengue virus exposure \cite{Dejnirattisai2016}. The complex ecology of pathogen communities therefore has potentially important implications for the control of infectious diseases and public health preparedness.

While some interactions between pathogens have been suggested by anecdotal epidemiological observations, statistical support for their occurrence and impact on the pathogen population dynamics is lacking. This is owing in part to a paucity of appropriate long-term time series data for many disease systems that describe infection frequencies for multiple pathogens simultaneously. Moreover, although several statistical methods are available for multivariately analysing health-related time series data (including cross-correlation, generalized linear models, wavelet decomposition and spectral analyses \cite{Dominici2002, Bishop1977, Pascual2000, Cazelles2007}) retrospectively inferring non-independent patterns among multiple time series has not been the primary focus.

Bayesian disease mapping models represent a class of regression model that has received much attention in recent years for the analysis of spatial distributions of incidence data routinely collected by public health bodies \cite{Lawson01082016, IntroDM}. These models are typically applied to incidence data to estimate spatial patterns of disease risk over a geographical region – with several models proposed to capture spatial autocorrelations \cite{Pascual2000} using conditional autoregressive priors \cite{RR4NoInteractions, Comparison1}.  While extensions to disease mapping models have been made to include temporal patterns \cite{SpaceTime2, RR4NoInteractions} and space-time interactions \cite{SpaceTime1, RR4}, most disease mapping applications focus on spatial structures \cite{AutoTime} with temporal dependencies in disease incidence often being overlooked \cite{Yorshire, Outbreak2}. However, multivariate forms of disease mapping models provide a suitable framework for estimating temporal dependencies between pathogens as they naturally incorporate a between-disease (or pathogen) covariance matrix \cite{Review1}.

Modelling multiple pathogens simultaneously allows assessment of specific patterns and non-independencies of infections risk among
different pathogens. By estimating the between-pathogen covariance matrix, we aimed to develop a statistical tool that readily enables the joint estimation of pathogen dependencies on the temporal dimension and distinguishes simple correlations from genuine pathogen-pathogen interactions. To validate this method we used two simulations studies and diagnostic data on five respiratory viruses (adenovirus [AdV], coronavirus [Cov], human metapneumovirus [MPV], influenza B virus [IBV] and respiratory syncytial virus [RSV]) from the patient population of a major urban UK population (Glasgow, United Kingdom) over a period of nine years. We chose this particular group of pathogens because i) respiratory viruses are obligate intracellular pathogens that have a strong predilection for the cells of the respiratory tract (i.e. they share the same ecological niche); ii) contemporary diagnostic tests based on multiplex real-time PCR (qPCR) technology allow the detection of multiple respiratory viruses; and iii) multiplex qPCR was routinely used to diagnose respiratory viruses in the patient population of Glasgow from 2005.

%\subsubsection*{Interactions between respiratory viruses}

\section*{Materials and methods}

\subsection*{Respiratory virus infection time series data  }
Our dataset derives from clinical samples tested for respiratory viruses by the West of Scotland Specialist Virology Center (WoSSVC) for Greater Glasgow and Clyde Health Board between January 2005 and December 2013. Each sample was tested by multiplex real-time RT-PCR and test results (virus positive or negative) were available for five groups of respiratory viruses: adenovirus [AdV]; coronavirus [CoV]; human metapneumovirus [MPV]; influenza B virus [IBV]; and respiratory syncytial virus [RSV]  \cite{Gunson2005}.  Sampling date, patient age, patient gender, the sample origin (hospital or general practice submission that we used as a proxy for infection severity) were recorded.  Multiple samples from the same patient received within a 30-day period were aggregated into a single episode of respiratory illness resulting in 28,647 patient episodes. A patient was considered virus-positive during an episode if at least one clinical sample was positive during the 30-day window.  We refer the reader to Nickbakhsh et al. \cite{Sema} for a full description of these data.

Whilst data are available at the individual level, we are predominantly interested in estimating non-independent patterns in temporal patterns between the five viruses at the population level.  Therefore, for each virus, data were aggregated into monthly infection counts across the time frame of this study.

\begin{figure}[!h]
\centering
\includegraphics[width=10cm,height=10cm]{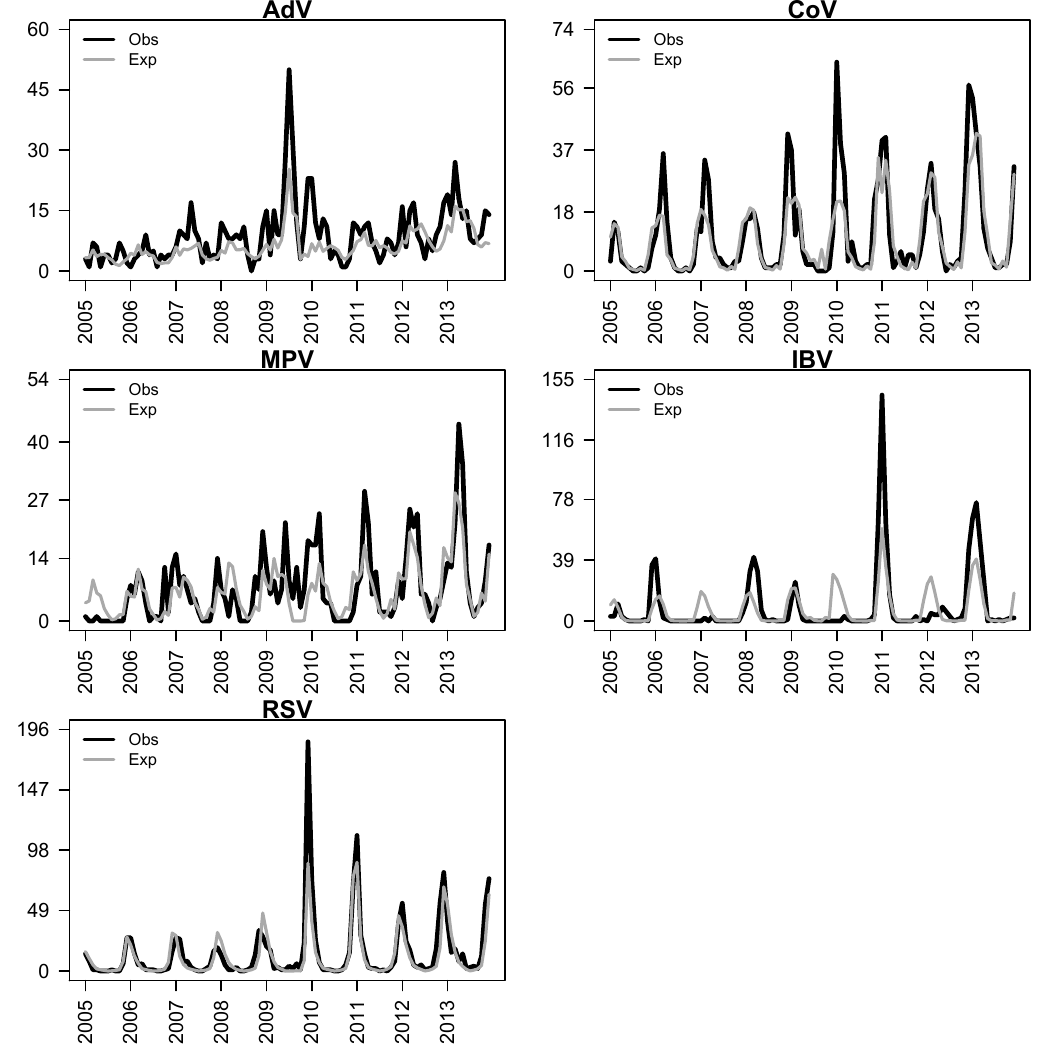}
\caption{{\bf Observed versus expected counts.} Observed (Obs) and expected (Exp) counts of the five groups of respiratory viruses between 2005 and 2013 (black and grey lines respectively). A full description of the estimated expected counts is given in the expected count section.}
\label{fig1}
\end{figure}

From the 28,647 patient episodes, 4,759 were positive to at least one virus group and detection was most common in children aged between 1 and 5 years \cite{Sema}. Detection of any virus in a given clinical sample was most common in December and least common in August. We observed differing patterns between the five viruses (Fig~\ref{fig1}, black lines). IBV, RSV and CoV were more prevalent in winter,  AdV was generally less common with a slight increase prevalence in spring and MPV shifts from winter peaks to summer peaks after 2010 \cite{Sema}.

Relative risks identify time points where observed counts are higher or lower than expected, with expected counts accounting for expected seasonality and risk factors associated with respiratory infection \cite{Sema}. Conventionally, relative risk would measure the risk of infection, if exposed to a risk factor, relative to the risk if unexposed. In this context, relative risk measures the risk of infection of a virus relative to an estimated expected risk.   Thus the relative risks measure the excess risk of viral infection that cannot be explained by seasonality or patient demographics.  Therefore, by inferring dependencies between viral species in terms of excess risks, we can directly infer viral interactions.  We provide a full description of the estimated expected risks in the expected count section.

\subsection*{Multivariate Spatio-temporal model} \label{MSTM}
Conditional autoregressive models are extensively used in the analysis of spatial data to model the relative risk of a virus or more generally a disease \cite{BDM, CAR1}.  The class of Bayesian model typically used in this context is given by	
\begin{equation}
\begin{split}
Y_{i}|E_{i},RR_{i} &\sim \mbox{Poisson}(E_{i}RR_{i}) \\
\log(RR_{i}) &= \alpha + \phi_{i} \nonumber
\end{split}
\end{equation}
where $Y_{i}$, $E_{i}$ and $RR_{i}$ are the observed count, expected count, derived from available patient demographic data (refer to expected counts section),  and relative risk for some index $i$ (for example, location or time interval) \cite{Comparison1} and $\phi=\{\phi_{1}, \ldots, \phi_{I}\}$ spatial random effects modelled jointly through a conditional autoregressive (CAR) distribution \cite{MvDM1}
\begin{eqnarray*}
\phi &\sim& \mbox{MVN}(0, (\tau(D-\lambda W))^{-1}).
\end{eqnarray*}
Matrix $W$ is a proximity matrix, $\lambda$ a smoothing parameter, $\tau$ a measure of precision and $D$ a diagonal matrix such that $D_{i}=\sum_{i'}{W_{ii'}}$.

Extending this model to multiple viruses, or more generally multiple pathogens, then
\begin{equation}
\begin{split}
Y_{iv}|E_{iv},RR_{iv} &\sim \mbox{Poisson}(E_{iv}RR_{iv}) \\
\log\frac{}{}(RR_{iv}) &= \alpha_{v} + \phi_{iv} \nonumber
\end{split}
\end{equation}
where $Y_{iv}$, $E_{iv}$ and $RR_{iv}$ are the observed count, expected count and relative risk of virus $v$ and $\alpha_{v}$ a virus specific intercept term. A multivariate CAR (MCAR) distribution can jointly model $\phi$ by incorporating a between virus covariance matrix $\Lambda^{-1}$ of dimension $V \times V$ (where $V$ is the total number of viruses):
\begin{eqnarray*}
\phi &\sim& \mbox{MVN}(0, [\Omega \otimes \Lambda ]^{-1}).
\end{eqnarray*}
In this case, $\Omega=D-\lambda W$, $\phi=\{\phi_{.1}, \ldots, \phi_{.V}\}$ and $\phi_{.v}=\{\phi_{1v},\ldots, \phi_{Iv}\}$ \cite{MvDM3, MvDM2}.

Temporal autocorrelations may be induced in this model, at time point $j$, through the conditional expectation of $\phi_{j}|\phi_{j-1}$
\begin{eqnarray*}
\phi_{j}|\phi_{j-1} &\sim& \mbox{MVN}(s\phi_{j-1}, [\Omega \otimes \Lambda ]^{-1}).
\end{eqnarray*}
The parameter $s$ controls the level of temporal autocorrelation such that $s=0$ implies no autocorrelation whereas $s=1$ is equivalent to a first order random walk \cite{SpaceTime1}. Typically, where temporal autocorrelations are modelled through the conditional expectation, spatial autocorrelations are modelled through the precision matrix \cite{SpaceTime1}.

\subsection*{Modelling multivariate time series data}
 We aim to model monthly time series count data from multiple viruses simultaneously over a nine year period. We index over monthly time intervals and so monthly autocorrelations are modelled in terms of the precision matrix and yearly autocorrelations are modelled in terms of the conditional expectation in a similar fashion to the multivariate spatial-temporal model detailed above.  The observed count of virus $v$ in month $m$ of year $t$, $Y_{mtv}$ is modelled in terms of the expected count $E_{mtv}$ and relative risk $RR_{mtv}$:
\begin{equation}%\tag{Model 4}
\begin{split}
Y_{mtv}|E_{mtv},RR_{mtv} &\sim \mbox{Poisson}(E_{mtv}RR_{mtv}) \\
\log(RR_{mtv}) &= \alpha_{v} + \phi_{mtv} \nonumber%\label{eqn5}
\end{split}
\end{equation}
with $\alpha_{v}$ an intercept term specific to virus $v$ and $\phi_{.t.}=\{\phi_{.t1}, \ldots, \phi_{.tV}\}$ a vector of random effects modelled conditionally through a MCAR prior
\begin{eqnarray*}
\phi_{.t.}|\phi_{.t-1.} &\sim& \mbox{MVN}(s_{v}\phi_{.t-1.}, [\Omega \otimes \Lambda]^{-1}).
\end{eqnarray*}
This parameterisation of a MCAR model captures both the seasonal trends of each virus via $\Omega$ and long-term temporal trends via $s_{1}, \ldots, s_{V}$. The conditional expectation of $\phi_{.t.}$ depends on the previous year $\phi_{.t-1.}$, capturing long term temporal tends.  By allowing dependencies between neighbouring months, we account for seasonality in viral infection frequencies.

\subsection*{Inferring viral interactions}
We focus primarily on the estimation of covariance matrix $\Lambda^{-1}$ in order to infer potential temporal dependencies.  By formally testing which off-diagonal entries of $\widehat{\Lambda}^{-1}$ are significantly different from zero, we can explicitly provide statistical support for viral interactions.

\subsubsection*{MCAR prior specification} \label{section:precision}
The covariance structure of the MCAR distribution used to model random seasonal-temporal effects is the Kronecker product of precision matrices $\Omega$ and $\Lambda$.

The between-virus precision matrix $\Lambda$ accounts for dependencies between viral relative risks in terms of monthly trends.  Wishart priors can be used for unstructured precision matrices such as $\Lambda$ \cite{WishartPrior}, however, we employed a modified Cholesky decomposition to estimate covariance matrix $\Lambda^{-1}$:
\begin{eqnarray*}
\Lambda^{-1} &=& \Sigma\Gamma\Gamma^{T}\Sigma
\end{eqnarray*}
where $\Sigma$ was a diagonal matrix with elements proportional to viral standard deviations and $\Gamma$ a lower triangular matrix relating to viral correlations \cite{MChol1}. This parameterisation ensured the positive-definiteness of $\Lambda^{-1}$, although we note that other parameterisations are available \cite{MChol2}.

Matrix $\Omega$ captures seasonal trends in terms of monthly dependencies defined through a proximity matrix $W$.  We will consider two possible constructions of $W$.

\subsubsection*{Neighbourhood structure}
Assuming neighbouring months are more similar than distance months, $W$ can be defined such that $w_{ij}=1$ if months $i$ and $j$ are neighbouring months and $w_{ij}=0$ if months $i$ and $j$ are not neighbouring months.  In this paper, neighbours were fixed as the previous and subsequent three months.  Taking a neighbourhood approach, we set
\begin{eqnarray*}
\Omega_{neigh} &=& D - \lambda W_{neigh}
\end{eqnarray*}
where $\lambda$ is a smoothing parameter and $D$ a $12 \times 12$ diagonal matrix with $D_{i}=\sum_{j}{w_{neigh_{ij}}}$. The total number of nearest neighbours of month $i$ \cite{MvDM3,CARSAR}.

\subsubsection*{Autoregressive structure}
Under this construction, $W$ was defined through an autogressive process and the corresponding matrix denoted by $W_{auto}$. We set the $ij$th entry of $W_{auto}$ ($i \neq j$) to be $\rho^{d_{ij}}$ with $d_{ij}$ the distance between months $i$ and $j$ and $\rho$ a temporal correlation parameter satisfying $\rho<1$.  We defined distance as the number of months between $i$ and $j$.

Taking an autoregressive approach, we set
\begin{eqnarray*}
\Omega_{auto} &=& D - \lambda W_{auto}
\end{eqnarray*}
with $D$ a diagonal matrix with $D_{i}=\sum_{j}{w_{auto_{ij}}}.$  We note that these formulations can easily be extended to other MCAR structures \cite{MvDM3,CARpriors}.

\subsubsection*{Expected counts} \label{section:expected}
We required expected counts of each virus at each time point in this study. Since individual level data were available, a series of logistic regressions were used to estimate the probability of testing positive for a virus at a given time point.  For month of the year $m$, the log odds of virus $v$, logit$(p_{mv})$, was estimated through fixed effects of age, sex and severity (estimated by hospital or general practice submission) and a yearly random effect.  The standardised probability of virus $v$ in month $m$, $p^{s}_{mv}$,  was estimated as
\begin{eqnarray*}
\widehat{\mbox{p}}^{s}_{mv} &=& \sum_{a,s,l,t}{\frac{N_{aslt}\widehat{p}_{mv_{aslt}}}{N_{mv}}}.
\end{eqnarray*}
where $N_{aslt}$ was the number of people of age $a$, sex $s$ and infection severity $l$ in year $t$; $\widehat{p}_{mv_{aslt}}$ the estimated probability of a person of age $a$, sex $s$ with infection severity $l$ in year $t$ testing positive for virus $v$ in month $m$; and $N_{mv}$ the number of swabs tested for virus $v$ in month $m$.  The estimated probabilities of each virus in each month are therefore standardised for age, sex and severity and account for yearly differences in circulation.

The expected count for virus $v$ in month $m$ of year $t$ was then
\begin{eqnarray*}
E_{mtv}=N_{mtv}\widehat{\mbox{p}}^{s}_{mv}
\end{eqnarray*}
with $N_{mtv}$ the number of of swabs tested for virus $v$ in month $m$ in year $t$.

\subsubsection*{Full model} \label{section:fullmodel}
\begin{figure}[!t]
\centering\includegraphics[width=10cm,height=13cm]{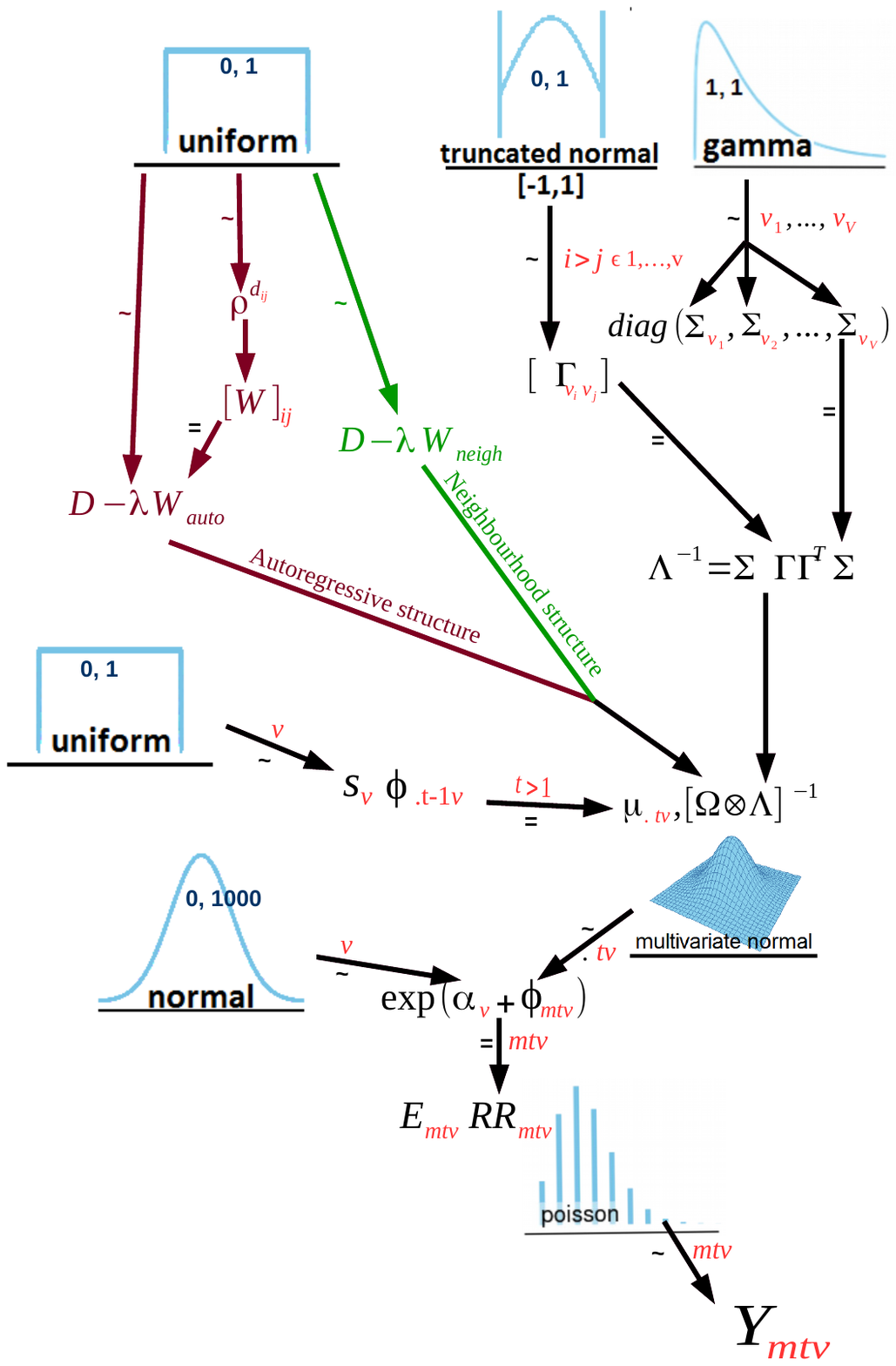}
\caption{{\bf Full model.} Full model used to estimate pairwise relative risk covariances. The diagram should be read from the bottom (starting with $Y_{mtv}$) to the top.  All prior choices have been fully specified.  Numbers in dark blue indicate hyperparameter choices, for instance, mean and variance in the normal distribution, lower and upper bound in the uniform distributions and shape and rate in the gamma distribution. Numbers in red indicate all relevant subscripts month $m=1,\ldots,12$, year $t=1,\ldots, 9$ and virus $v=1,\ldots,5$. Green areas refer to the neighbourhood structure and maroon areas refer to the autoregressive structure.}
\label{fig_fullmodel}
\end{figure}

Let $Y_{mtv}$ denote the observed count of virus $v$ during the $m$th month of year $t$, $V$ the total number of viruses and $T$ the number of years. We modelled $Y_{mtv}$ in terms of the expected count $E_{mtv}$ and relative risk $RR_{mtv}$ and used a Cholesky decomposition to estimate covariance matrix $\Lambda^{-1}$ \cite{DBDA} (Fig~\ref{fig_fullmodel}).

This model was implemented in jags \cite{jags} using the R2jags package \cite{R2jags} in R \cite{R}.  All results are averaged across five independent chains.  In each chain, we took 50,000 thinned draws across 500,000 iterations after a burn-in period of 300,000 iterations.  Our R code used to model these data and an example of simulated are available upon request to the first author.  We note that the multivariate intrinsic Gaussian CAR prior distribution is fully specified in GeoBUGS \cite{GeoBUGS}.  However, our approach allows for other parameterisations of the MCAR distribution providing more flexibility in separating monthly and yearly temporal dependencies.

\subsection*{Simulation Studies}
The specific aim of this paper was to estimate the between-virus covariance matrix $\Lambda^{-1}$.  We show the validity of our proposed model (Fig~\ref{fig_fullmodel}) in modelling multivariate time series data through a three-virus simulation and that this model can accurately and precisely estimate $\Lambda^{-1}$ through a five-virus simulation.

\subsubsection*{Three-virus example}
\begin{figure}[!h]
    \centering
    \begin{subfigure}[h]{0.4\textwidth}\label{fig_3virus_setup_a}
    \centering
        \includegraphics[height=5cm,width=6cm]{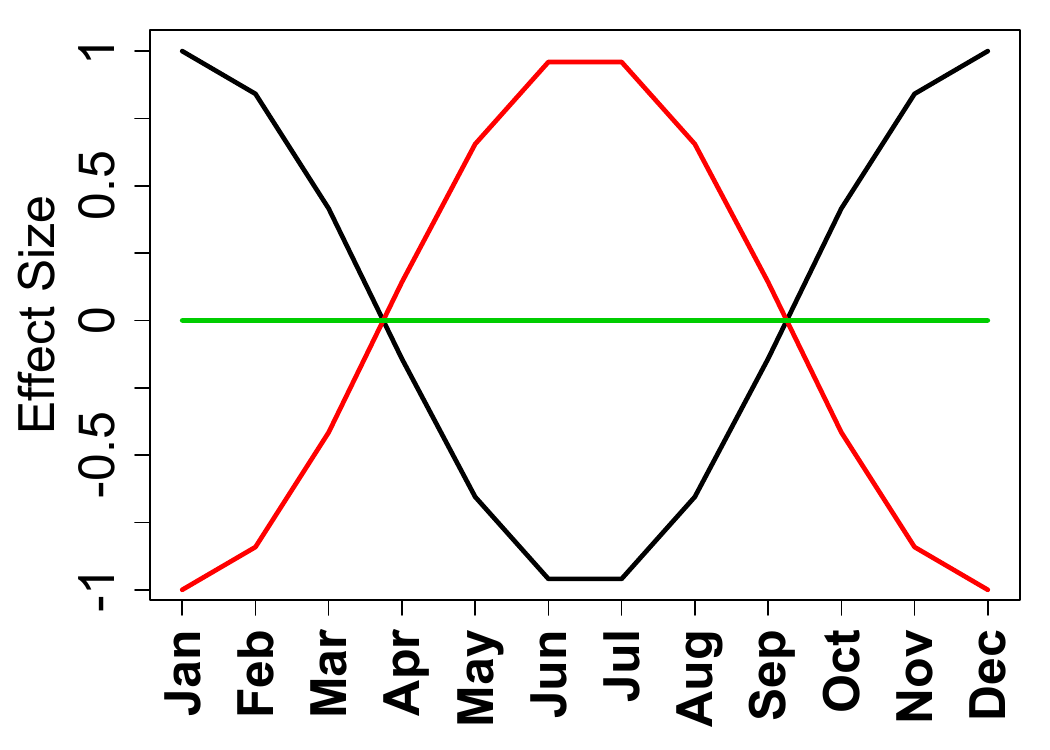}
        \caption{}
    \end{subfigure}
    \hspace{1.25cm}
    \begin{subfigure}[h]{0.4\textwidth}\label{fig_3virus_setup_b}
    \centering
    \includegraphics[height=5cm,width=6cm]{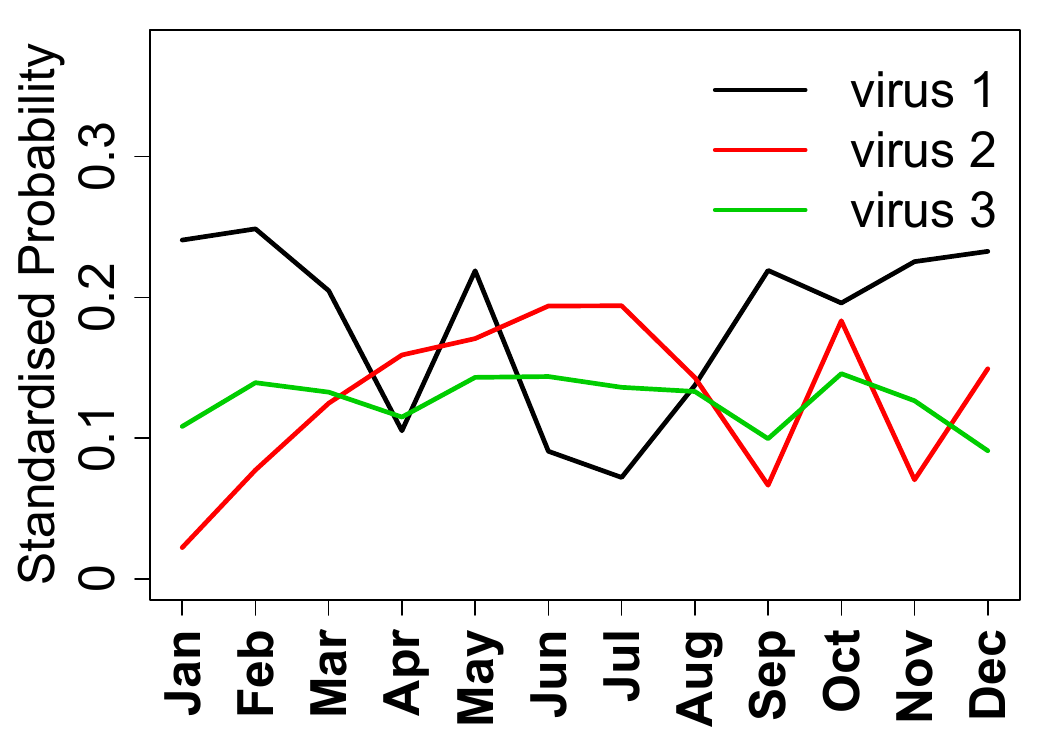}
        \caption{}
    \end{subfigure}
   % \hspace{-1cm}
    \begin{subfigure}[h]{1\textwidth}\label{fig_3virus_setup_c}
    \centering
        \includegraphics[height=5cm,width=14cm]{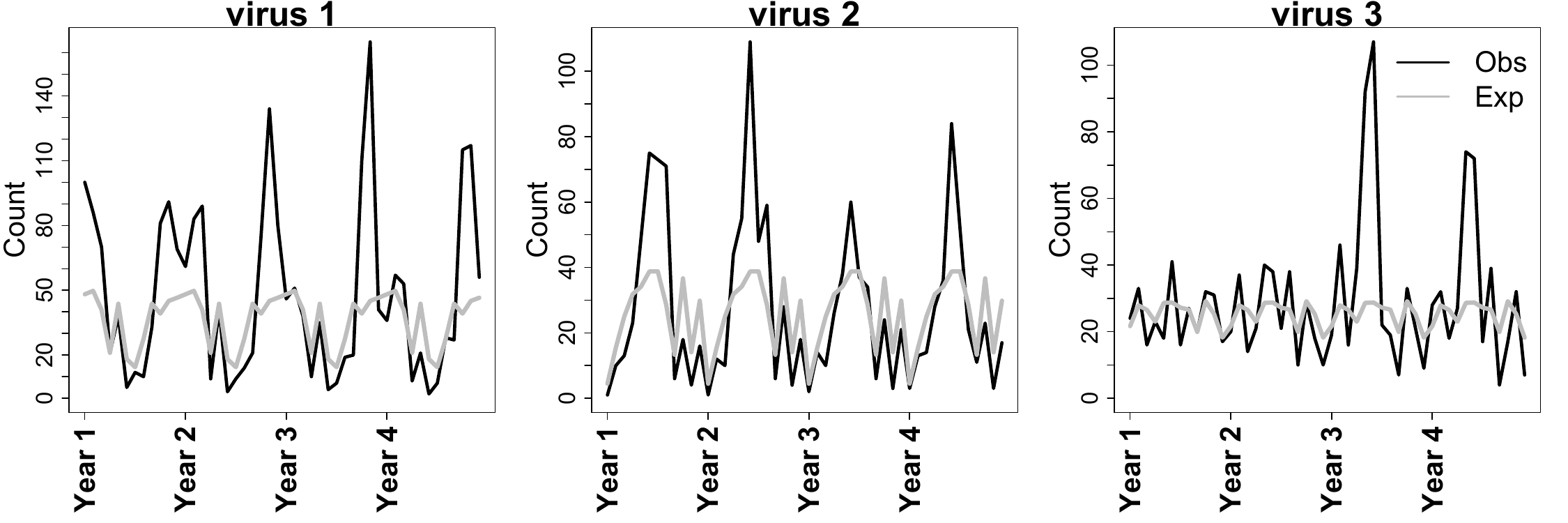}
        \caption{}
    \end{subfigure}
  \caption{{\bf Three virus simulation.} (a) Monthly effect sizes of virus 1, black, virus 2, red, and virus 3, green. (b) estimated monthly standardised probabilities of virus 1, black, virus 2, red, and virus 3, green. (c) simulated expected (grey lines) and observed (black lines) counts of virus 1, virus 2 and virus 3 in three virus simulation.}
 \label{fig_3virus_setup}
\end{figure}

We first present an analysis of simulated data for three viruses; virus 1, virus 2 and virus 3. Seasonal effects were assigned such that virus 1 peaked in winter, virus 2 peaked in summer and virus 3 had no seasonal pattern (Fig ~\ref{fig_3virus_setup}).  Virus 1 and virus 2 were negatively correlated ($\Lambda_{12}=-0.5$) and both viruses were independent of virus 3 ($\Lambda_{13}=\Lambda_{23}=0$).

%\begin{table}[!h]
%\caption{True values of matrix $\Lambda$ from simulation study and estimated 95\% credible intervals from full model using the neighbourhood ($W_{neigh}$, true model) and autoregressive ($W_{auto}$) structures.
%\label{table1}}
%{\begin{tabular}{cccc}
%  \hline
  % after \\: \hline or \cline{col1-col2} \cline{col3-col4} ...
%  \mbox{parameter} & \mbox{true value} & \mbox{95\% credible interval} ($W_{neigh}$) & \mbox{95\% credible interval} ($W_{auto}$) \\ \hline
  %$\Lambda^{-1}_{1,1}$ & 1   & $(0.56, 1.54)$ \\
%  $\Lambda^{-1}_{1,2}$ & -0.5 & $(-0.89, -0.30)$ & $(-0.72,-0.36)$ \\
%  $\Lambda^{-1}_{1,3}$ & 0   & $(-0.39, 0.27)$  & $(-0.40, 0.32)$ \\
  %$\Lambda^{-1}_{2,2}$ & 1   & $(0.48, 1.41)$ \\
%  $\Lambda^{-1}_{2,3}$ & 0   & $(-0.32, 0.35)$ & $(-0.37, 0.39)$ \\
  %$\Lambda^{-1}_{3,3}$ & 1   & $(0.37, 1.23)$ \\
%  \hline
%\end{tabular}}
%\end{table}

The probabilities and expected counts of each virus in each month were estimated using the method described in \nameref{section:expected}. Individual level data were simulated in order to reflect the virological diagnostic data.  We simulated 200 samples per month per virus over a four year period.  For each sample, an age, sex and severity were drawn from the observed virological diagnostic data distributions \cite{Sema}. Regression coefficients used to estimate the probability of each virus were drawn such that $\beta_{intercept}=0$, $\beta_{age} \sim N(0,0.1)$, $\beta_{gender} \sim N(0,0.1)$ and $\beta_{severity} \sim N(0,0.1)$ and standardised probabilities of each virus in each month were estimated (example shown in Fig~\ref{fig_3virus_setup}).  Expected counts were taken as the product of the standardised probabilities and the number of samples (i.e. 200) (example shown in Fig~\ref{fig_3virus_setup}c, grey lines).

Random effects $\phi$ were drawn from multivariate normal distributions with yearly smoothing parameters and monthly smoothing parameter ($s_{1}, s_{2}$, $s_{3}$ and $\lambda$) set at 0.5.  Seasonal dependencies were simulated through neighbourhood matrix $W_{neigh}$ defined in \nameref{section:precision}.  We set virus specific intercept terms $\alpha_{1}=\alpha_{2}=\alpha_{3}=0$ and calculated the relative risks of each virus at each time point. Finally, observed counts were taken as the product of relative risks and expected counts (example shown in Fig~\ref{fig_3virus_setup}c, black lines).

We fitted both models (Fig~\ref{fig_fullmodel}, neighbourhood and autoregressive structure) to data simulated only through the neighbourhood structure and estimated higher posterior density intervals (HPDI) for each covariance parameter ($\hat{\Lambda}_{12}$,$\hat{\Lambda}_{13}$ and $\hat{\Lambda}_{23}$). Posterior probabilities were then estimated to assess the probability of zero being included in each interval, synonymous to  Bayesian p-values defined in terms of lower tail posterior probabilities \cite{pval1,pval2}. Covariance parameters with a posterior probability less than 0.05 were deemed different from zero \cite{pval1}. In order to control for multiple comparisons, covariance parameters with an adjusted probability, controlling the false discovery rate \cite{pval1,MCP}, less than 0.05 were deemed different from zero and used as support for a significant covariance between the corresponding viruses.

Generally, significant negative covariances were estimated between virus 1 and virus 2 and non-significant covariances between virus 1 and virus 3 and between virus 2 and virus 3 using both the neighbourhood and autoregressive constructions. %(Table~\ref{table1}, 95\% credible intervals).

Under the neighbourhood structure, the probability of correctly identifying a non-zero covariance was 0.98 (Table ~\ref{tab1}, $\hat{\Lambda}_{12}$ $1- \mbox{Pr}\{0 \in HPDI\}$).  However, the probabilities of falsely identifying a non-zero covariance were 0.13 and 0.1 (Table ~\ref{tab1}, $\hat{\Lambda}_{13}$ and $\hat{\Lambda}_{23}$ $1- \mbox{Pr}\{0 \in HPDI\}$).

Under the autoregressive structure, the probability of correctly identifying a non-zero covariance was 0.87 (Table~\ref{tab1}, $\hat{\Lambda}_{12}$ $1- \mbox{Pr}\{0 \in HPDI\}$).  However, the probabilities of falsely identifying a non-zero covariance were 0.07 and 0.1 (Table~\ref{tab1}, $\hat{\Lambda}_{13}$ and $\hat{\Lambda}_{23}$ $1- \mbox{Pr}\{0 \in HPDI\}$).

\begin{table}
%\begin{adjustwidth}{-2cm}{}
\caption{{\bf Type 1 error and power.} Higher posterior density intervals for covariance parameters $\Lambda_{12}$,$\Lambda_{13}$ and $\Lambda_{23}$ from simulated data estimated under the  neighbourhood structure and autoregressive structure before and after a correction for multiple comparisons (mcc).}
\label{tab1}
{\begin{tabular}{lllll}
 \hline
 % after \\: \hline or \cline{col1-col2} \cline{col3-col4} ...
 &Neighbourhood & &Autoregressive&\\
 & $1- \mbox{Pr}\{0 \in HPDI\}$ & $1- \mbox{Pr}\{0 \in HPDI_{mcc}\}$ & $1- \mbox{Pr}\{0 \in HPDI\}$ & $1- \mbox{Pr}\{0 \in HPDI_{mcc}\}$ \\ \hline
$\hat{\Lambda}_{12}$ & 0.98 (0.95, 1.00) & 0.94 (0.89, 0.99) & 0.87 (0.81, 0.94) & 0.71 (0.62, 0.80) \\
$\hat{\Lambda}_{13}$ & 0.13 (0.06, 0.20) & 0.05 (0.01, 0.09) & 0.07 (0.02, 0.12) & 0.03 (0.00, 0.06) \\
$\hat{\Lambda}_{23}$ & 0.11 (0.04, 0.16) & 0.04 (0.00, 0.08) & 0.10 (0.04, 0.15) & 0.04 (0.001, 0.07) \\
\hline
\end{tabular}}
%\end{adjustwidth}
\end{table}

Correcting for multiple comparisons (mcc) reduced the probabilities of falsely identifying a non-zero covariance to 0.05 and 0.04 under the neighbourhood structure and 0.03 and 0.04 under the autoregessive structure (Table~\ref{tab1}, $\hat{\Lambda}_{13}$ and $\hat{\Lambda}_{23}$ $1- \mbox{Pr}\{0 \in HPDI_{mcc}\}$).

Power estimates under the neighbourhood structure were reduced to 0.94  and 0.71 under the autogressive structure (Table~\ref{tab1}, $\hat{\Lambda}_{12}$ $1- \mbox{Pr}\{0 \in HPDI_{mcc}\}$). A higher estimated power was expected under the neighbourhood structure since all data were simulated under the neighbourhood structure.

Reducing the covariance between virus 1 and virus 2 ($\Lambda_{12}=-0.25$) decreased the power from 0.94 (under the neighbourhood structure with the multiple comparisons correction) to 0.64 and a further reduction to $\Lambda_{12}=-0.1$ reduced the corresponding power to  0.22.  Therefore, as expected, this test has high power to detect strong associations but lacks the ability to detect an association between viruses with a small dependency.

%{{\color{red}{[I am running with smaller correlations ($\Lambda_{12}=-0.25 \mbox{ and } -0.1$) to see how power decreases for weaker associations.]}}

%,  and found little difference in fit between the neighbourhood construction.  (average DIC = 960.7) and the autoregressive construction (average DIC = 957.6).

%By taking the average absolute difference, $d$, between the true relative risks and the estimated relative risks using $W_{neigh}$ we found the estimated pattern of relative risks across the time frame of this simulation followed closely to the true relative risks (Fig~\ref{figSS_RR}).   We found very similar results using $W_{auto}$ (results not shown).

%\begin{figure}[!h]
%\centering
%\includegraphics[width=9cm,height=9cm]{3Virus_RR_redo2}
%\caption{{\bf Difference between true and estimated relative risks.}Absolute difference $d$ between the true relative risks used in simulated study and the estimated relative risk.  The maximum difference was 0.3.}
%\label{figSS_RR}
%\end{figure}

This simulation study illustrates our method can powerfully and accurately estimate the relative risk of each virus across the time frame of the simulation whilst controlling the familywise error rate and is an important new tool in modelling multivariate time series data.

\subsubsection*{Five-virus example}

\begin{figure}[!t]
\centering
\includegraphics[width=12cm,height=12cm]{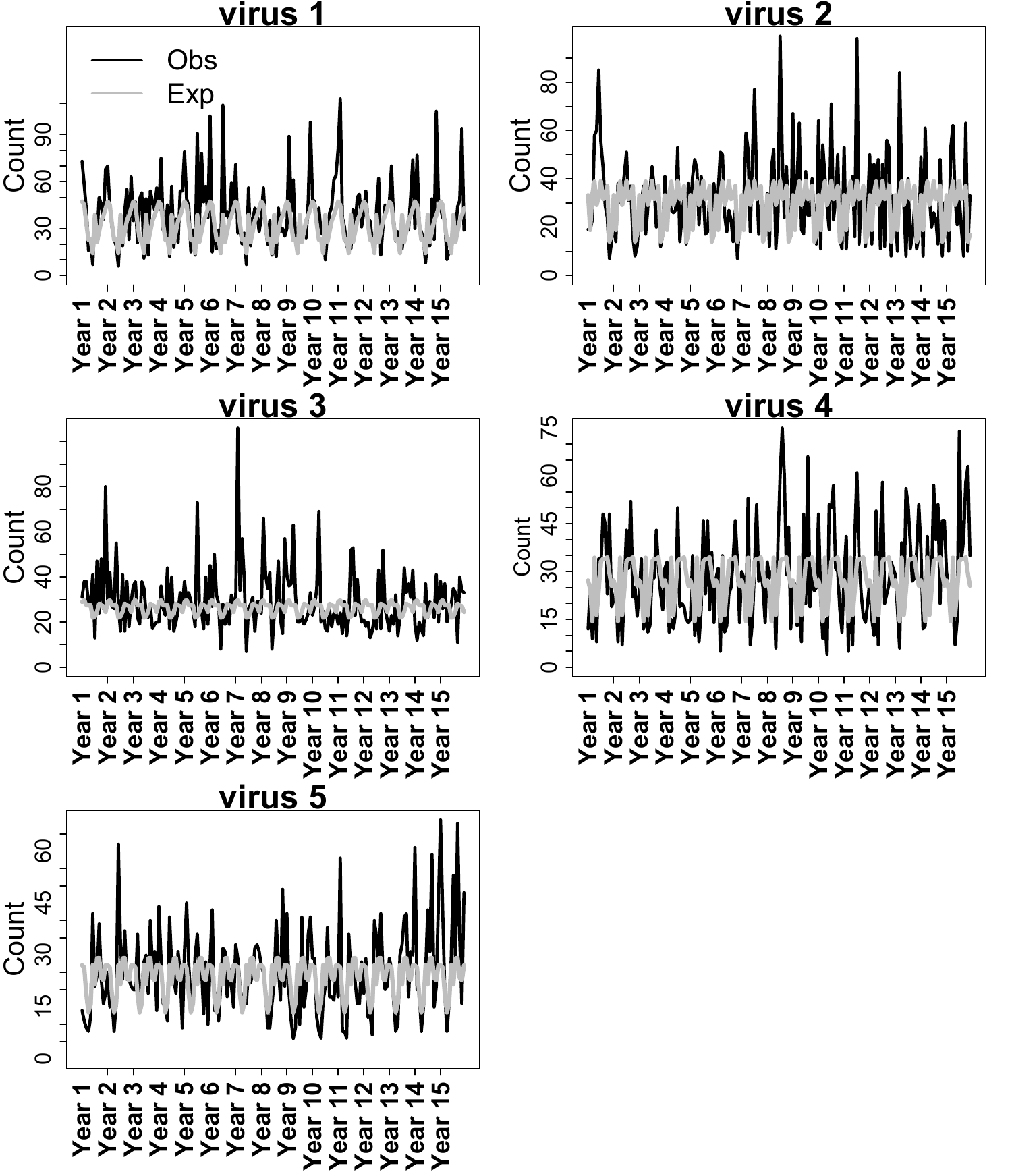}
\caption{{\bf Five virus simulation.} Simulated observed (black lines) and expected (grey lines) virus counts from five viruses over fifteen years.}
\label{figSS2}
\end{figure}

To test the accuracy of the full model in estimating the between virus covariance matrix, we simulated data from five viruses over a 15 year period (Fig~\ref{figSS2}) with
$$\Lambda^{-1}=
\left(
  \begin{array}{ccccc}
    1 & -0.5 & 0 & 0 & 0 \\
    -0.5 & 1 & 0 & 0 & 0 \\
    0 & 0 & 1 & 0 & 0 \\
    0 & 0 & 0 & 1 & 0.5 \\
    0 & 0 & 0 & 0.5 & 1 \\
  \end{array}
\right).
$$

Since temporal variances within each virus were set to 1 (diagonal entries of $\Lambda*{-1}$), we refer to virus covariances within this section as correlations. Virus 1 and virus 2 were negatively correlated (such that when the prevalence of one virus increases, the prevalence of the other decreases), virus 4 and virus 5 were positively correlated and virus 3 circulated independently.

Data from the first three viruses were simulated under identical conditions to those described in the three-virus example.  Seasonal effects were assigned such that virus 4 and virus 5 peaked in autumn (Fig~\ref{figSS2}, observed counts). The probabilities and expected counts of virus 4 and virus 5 in each month were estimated using the method described in \nameref{section:expected}.  Likewise, regression coefficients were drawn such that $\beta_{intercept}=0$, $\beta_{age} \sim N(0,0.1)$, $\beta_{gender} \sim N(0,0.1)$ and $\beta_{severity} \sim N(0,0.1)$ .  Random effects $\phi$ were drawn from multivariate normal distributions with yearly smoothing parameters and monthly smoothing parameter ($s_{1}$, $s_{2}$, $s_{3}$, $s_{4}$, $s_{5}$ and $\lambda$) set as 0.5.  Seasonal dependencies were simulated through neighbourhood matrix $W_{neigh}$ defined in \nameref{section:precision}.  We set virus specific intercept terms $\alpha_{1}=\alpha_{2}=\alpha_{3}=\alpha_{4}=\alpha_{5}=0$ and calculated the relative risks of each virus at each time point.  Finally, observed counts were taken as the product of relative risks and expected counts (Fig~\ref{figSS2}, grey lines).

\begin{figure}[!t]
\centering
\includegraphics[width=7cm,height=7cm]{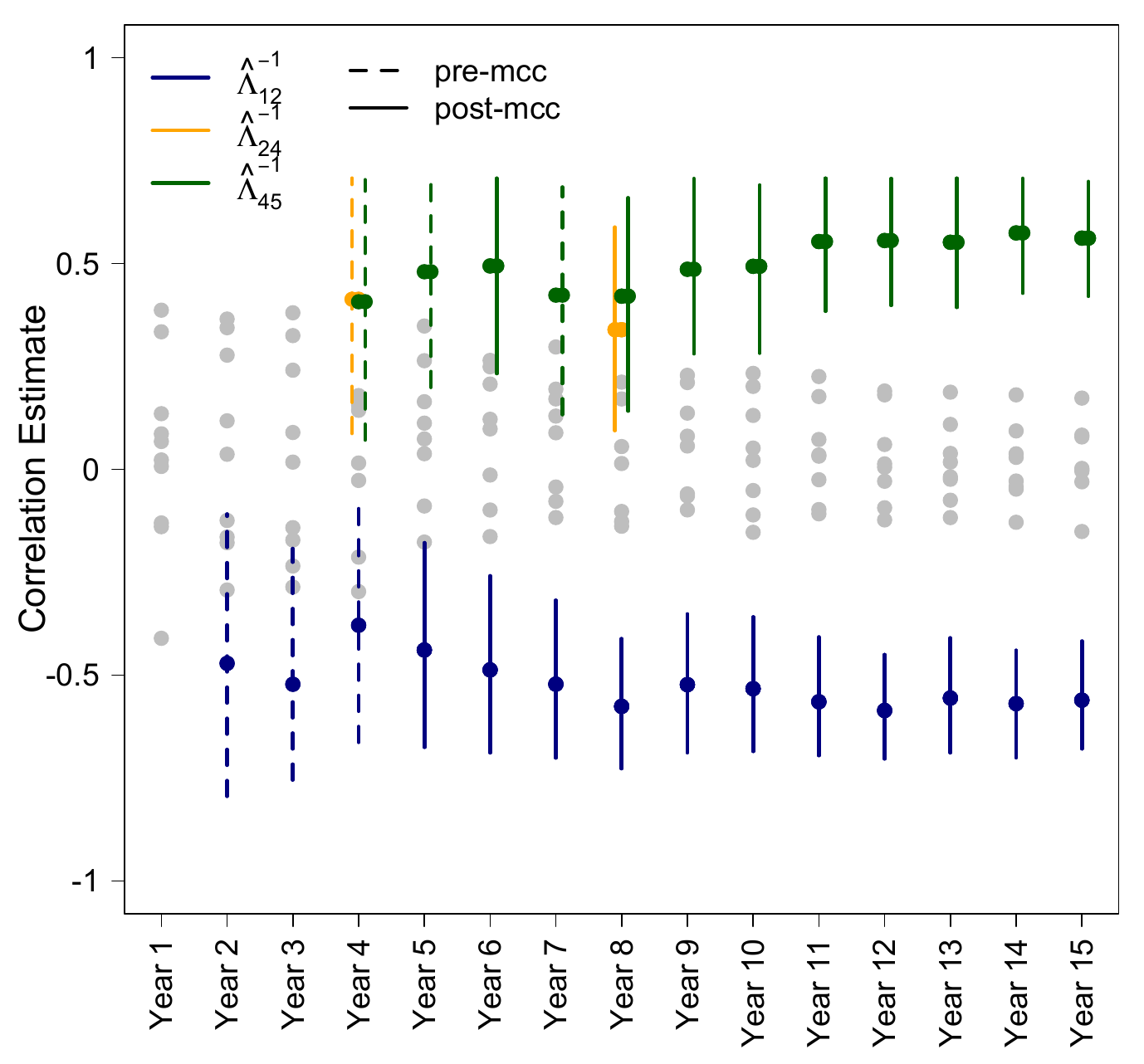}
\caption{{\bf Estimated correlations over 15 years.} Estimated correlation parameters significantly different from zero using data from year 1 through to year 15. Grey dots show point estimates of all correlation parameters at each time point.  Solid vertical lines show parameters significantly different from zero once multiple comparison correction (mcc) was applied. Dotted vertical lines show parameters only significantly different from zero before multiple comparison correction (pre-mcc). }
\label{figSS3}
\end{figure}

Using the full model with the neighbourhood construction, we initially estimated $\Lambda^{-1}$ using only the first year of data (Fig~\ref{figSS3}, Year 1).  We then combined the first two years of data to estimated $\Lambda^{-1}$ (Fig~\ref{figSS3}, Year 2).  This process was repeated, adding the current year's data at each iteration, until all fifteen years of data were combined to estimate $\Lambda^{-1}$.

For each pair of viruses, at each iteration, we assessed deviations of each  parameter from zero (Fig~\ref{figSS3}, pre-mcc) and applied the multiple comparison correction (Fig~\ref{figSS3}, post-mcc).  Using data from the Year 1, we estimated non-significant correlation between all pairs of viruses (Fig~\ref{figSS3}, Year 1).  Using data from Year 2 onwards, we found varying degrees of significance in the correlation between virus 1 and virus 2, $\Lambda^{-1}_{1,2}$, virus 2 and virus 4, $\Lambda^{-1}_{2,4}$, and virus 4 and virus 5, $\Lambda^{-1}_{4,5}$ (Fig~\ref{figSS3}, Year 2 onwards).

Following the multiple comparison correction, we found significant correlations between virus 1 and virus 2 ($\Lambda^{-1}_{1,2}$), virus 4 and virus 5 ($\Lambda^{-1}_{4,5}$) from year 9 onwards (Fig~\ref{figSS3}, solid navy and green lines) validating our model for long term time series data. We found no significant correlation between virus 2 and virus 4 from year 9 onwards. We found very similar results using $W_{auto}$ (results not shown).

This example shows this method can precisely and accurately estimate the between virus covariance matrix given long term time series data.

%PLOS does not support heading levels beyond the 3rd (no 4th level headings).
% Place tables after the first paragraph in which they are cited.
% Results and Discussion can be combined.
\section*{Results}

\subsection*{Application to virological diagnostic data}
We applied our model (Fig~\ref{fig_fullmodel}) to monthly infection count data from five respiratory viruses (AdV, Cov, MPV, IBV and RSV) across nine years in order to infer interactions between these viruses.

For comparison, we applied the full model assuming all five viruses to be independent by setting $\Lambda^{-1}=I_{5}$.  Under the neighbourhood structure, we found that allowing dependencies between viruses ($\Lambda^{-1}\neq I_{5}$) provided a better fit to these data (DIC=2795.6 compared to DIC=3583.8). However, the autoregressive structure with $\Lambda^{-1}\neq I_{5}$ minimised DIC (DIC=2686.4).

\subsubsection*{Estimating $RR$ and $\Lambda^{-1}$}
Expected counts were estimated for each virus as described in \nameref{section:expected} (Fig~\ref{fig1}, grey lines). The risk of AdV infection remained relatively high between 2005 to 2010 but decreased during the summer and autumn months of 2011, 2012 and 2013 . We found an increased risk of IBV infection during the autumn and winter periods of 2005/2006, 2010/2011 and 2012/2013 . During the second half of 2009, we found a heightened risk of RSV and MPV infections.  More generally, the risk of RSV infection peaked during late summer through to autumn from 2008 onwards whereas the risk of MPV infection shifted from winter, between 2005 and 2008, to summer, from 2011 onwards.
%(Fig~\ref{figRR_11}, RSV and MPV).
%(Fig~\ref{figRR_11}, AdV)
%(Fig~\ref{figRR_11}, IBV)
%\begin{figure}[!h]
%\centering
%\includegraphics[width=10cm,height=10cm]{"Figure7"}
%\caption{{\bf Estimated risks of five viruses.} Estimated excess risks of adenovirus [AdV], coronavirus [Cov], human metapneumovirus [MPV], influenza B virus [IBV] and human Respiratory syncytial virus [RSV] 2005 and 2013 that cannot be explained by seasonality or patient demographics.}
%\label{figRR_11}
%\end{figure}

Under the neighbourhood structure, we found a positive covariance between  RSV \& MPV and negative covariances between IBV \& MPV, Cov \& MPV and Adv \& IBV (Table~\ref{figCov_11}, $W_{neigh}$) whereas we found significant covariances between RSV \& MPV and a negative covariance between IBV \& AdV using the autoregressive structure (Table~\ref{figCov_11}, $W_{auto}$).  Under this construction, we found adjusted p-values for the covariances between IBV \& MPV and Cov \& MPV to be 0.075 and 0.073 respectively.

\begin{table}[!h]
\caption{Posterior density interval estimates of covariancess between adenovirus [AdV], coronavirus [Cov], human metapneumovirus [MPV], influenza B virus [IBV] and respiratory syncytial virus [RSV].  Covariances different from zero after multiple comparison correction are highlighted in bold.
\label{figCov_11}}

{\begin{tabular}{llcc}
 \hline
  % after \\: \hline or \cline{col1-col2} \cline{col3-col4} ...
      &      &  $W_{neigh}$ &  $W_{auto}$ \\ \hline
  Adv & Cov & (-0.27, 045) & (-0.31, 0.41)\\
      & MPV & (-0.35, 0.22) & (-0.35, 0.20)\\
      & IBV & \textbf{(-0.67, -0.16)} & \textbf{(-0.68, -0.15)}\\
      & RSV & (-0.37, 0.29) & (-0.32, 0.41)\\
  Cov & MPV & \textbf{(-0.66, -0.11) }  & (-0.66, -0.08)\\
      & IBV & (-0.23, 0.45) & (-0.18, 0.43)\\
      & RSV & (-0.28, 0.32) & (-0.32, 0.29)\\
  MPV & IBV & \textbf{(-0.66 -0.13)} & (-0.64, -0.07)\\
      & RSV & \textbf{(0.32, 0.71)} & \textbf{(0.18, 0.67)}\\
  IBV & RSV & (-0.51, 0.05) & (-0.54, 0.04)\\
  \hline
\end{tabular}}
\end{table}

\section*{Discussion}
Humans, animals and plants are exposed to a plethora of cocirculating pathogens, which creates frequent opportunity for ecological interactions between them. We adapted Bayesian multivariate disease mapping methods to provide a powerful statistical tool for inferring pathogen-pathogen interactions from diagnostic and/or surveillance time series data. Whilst standard multivariate disease mapping frameworks investigate the joint spatial distribution of multiple diseases coinfecting a population simultaneously, our method instead analyses the joint temporal distribution of multiple infections. Because multivariate disease mapping naturally incorporates a between-disease covariance matrix, these methods conveniently lend themselves to the inference of temporal signatures of pathogen-pathogen interactions when adapted to analyse temporal dependencies.

By applying our framework to extensive diagnostic data accrued over a nine-year period from a well defined patient population, our analysis provides evidence of epidemiological interactions among respiratory viruses. This is particularly important because acute respiratory infections are the most common cause of illness and mortality and are primarily attributed to a group of viruses that occupy a shared ecological niche in the respiratory tract. Although observational data \cite{Observational3,Observational2,interact2, Observational1} and univariate regression models \cite{Greer2009,interact2, interact3,ModellingReview1} indicate the potential for interactions among these common pathogens, limited evidence exists of their impact on epidemiological infection dynamics.

A recent study used wavelet decomposition to infer temporal correlations in the time series of a set of respiratory viruses \cite{Bhattacharyya2015 }. However, as those statistical methods do not enable different causal hypotheses to be tested, the authors were unable to determine whether pathogen-pathogen interactions were the consequence of cross-immunity or of other drivers of seasonal correlation. Under the autoregressive structure which provided a better fit to these data, our analysis provides robust evidence of a positive covariance between RSV and MPV and a negative covariance between IBV and AdV.

In summary, we adapted a multivariate disease mapping framework to retrospectively infer pathogen-pathogen interactions in a statistically robust fashion. Applying this approach to time series data of pathogens that co-circulate in a given population will lead to a better understanding of the joint epidemiological dynamics of diseases. It is anticipated that these inferences will enhance the understanding of linked pathogen dynamics and inform forecasting of disease incidence and improve public health preparedness. In addition, they will result in better ways to evaluate the impact of public health interventions thus aiding the design of measures to control infectious diseases.

%\section*{Supporting information}

% Include only the SI item label in the paragraph heading. Use the \nameref{label} command to cite SI items in the text.

\section*{Acknowledgments}
This work was funded by the Medical Research Council of the United Kingdom (Grant number MC\_UU\_12014/9).  We are grateful for support from National Science Foundation DEB1216040, BBSRC grants BB/K01126X/1, BB/L004070/1, BB/L018926/1, BB/N013336/1, BB/L004828/1, BB/H009302/1, BB/H009175/1, the Foods Standards Agency FS101055 and the Scottish Government Rural and Environment Science and Analytical Services Division, as part of the Centre of Expertise on Animal Disease Outbreaks (EPIC).  We thank Paul Johnson and Theo Pepler for their helpful comments on the manuscript.

%\nolinenumbers

% Either type in your references using
% \begin{thebibliography}{}
% \bibitem{}
% Text
% \end{thebibliography}
%
% or
%
% Compile your BiBTeX database using our plos2015.bst
% style file and paste the contents of your .bbl file
% here. See http://journals.plos.org/plosone/s/latex for
% step-by-step instructions.
%

\bibliographystyle{plos2015}
\bibliography{smmr2}

%\begin{thebibliography}{10}

%\bibitem{bib1}
%Conant GC, Wolfe KH.
%\newblock {{T}urning a hobby into a job: how duplicated genes find new
%  functions}.
%\newblock Nat Rev Genet. 2008 Dec;9(12):938--950.

%\bibitem{bib2}
%Ohno S.
%\newblock Evolution by gene duplication.
%\newblock London: George Alien \& Unwin Ltd. Berlin, Heidelberg and New York:
%  Springer-Verlag.; 1970.

%\bibitem{bib3}
%Magwire MM, Bayer F, Webster CL, Cao C, Jiggins FM.
%\newblock {{S}uccessive increases in the resistance of {D}rosophila to viral
%  infection through a transposon insertion followed by a {D}uplication}.
%\newblock PLoS Genet. 2011 Oct;7(10):e1002337.

%\end{thebibliography}

\end{document}